\documentclass[prb,twocolumn,showpacs,preprintnumbers,amsmath,amssymb]{revtex4}

\usepackage{graphicx}
\usepackage{dcolumn}
\usepackage{bm}

\begin{document}

\preprint{APS/123-QED}

\title{High--field NMR of the quasi--1D antiferromagnet LiCuVO$_4$}

\author{N. B\"{u}ttgen$^1$\email{norbert.buettgen@physik.uni-augsburg.de}, P. Kuhns$^2$, A. Prokofiev$^3$, A.P. Reyes$^2$, and L.E. Svistov$^4$\email{svistov@kapitza.ras.ru} }
\affiliation{$^1$ Center for Electronic Correlations and Magnetism
EKM, Experimentalphysik V, Universit\"{a}t Augsburg, D--86135
Augsburg, Germany \\
$^2$ National High Magnetic Field Laboratory, Tallahassee, FL 32310, USA\\
$^3$ Institut f\"{u}r Festk\"{o}rperphysik Technische
Universit\"{a}t Wien, A--1040 Wien, Austria\\
$^4$ P.L.Kapitza Institute for Physical Problems RAS, 119334 Moscow, Russia\\}

\date{\today}

\begin{abstract}
We report on NMR studies of the quasi one--dimensional (1D)
antiferromagnetic $S=1/2$ chain cuprate LiCuVO$_4$ in magnetic
fields $H$ up to $\mu_0H$ = 30 T ($\approx 70$~\% of the saturation
field $H_{\rm sat}$). NMR spectra in fields higher than $H_{\rm c2}$
($\mu_0H_{\rm c2} \approx 7.5$~T) and temperatures $T<T_{\rm N}$ can
be described within the model of a spin--modulated phase in which
the magnetic moments are aligned parallel to the applied field $H$
and their values alternate sinusoidally along the magnetic chains.
Based on theoretical concepts about magnetically frustrated 1D
chains, the field dependence of the modulation strength of the
magnetic structure is deduced from our experiments. Relaxation time
$T_2$ measurements of the $^{51}$V nuclei show that $T_2$ depends on
the particular position of the probing $^{51}$V nucleus with respect
to the magnetic copper moments within the 1D chains: the largest
$T_2$ value is observed for the vanadium nuclei which are very next
to the magnetic Cu$^{2+ }$ ion with largest ordered magnetic moment.
This observation is in agreement with the expectation for the
spin--modulated magnetic structure. The $(H,T)$ magnetic phase
diagram of LiCuVO$_4$ is discussed.
\end{abstract}

\maketitle

\section{INTRODUCTION}

Magnetic frustration in quantum--spin chains ($S=1/2$) with
competing nearest neighbor (NN) and next--nearest neighbor (NNN)
exchange interactions yields nontrivial magnetic order which attract
much attention from the theoretical point of
view.\cite{Chubukov_91,Maeshima_05,Kolezhuk_05,Heidrich_06,Kecke_07,Hikihara_08,
Sudan_09,Heidrich_09,Ueda_09,Zhitomirsky_10,Nishimoto_10} The
conventional magnetic order with non--zero mean value of the
magnetization at the magnetic ions is forbidden in 1D chains because
of strong spin fluctuations even at zero temperature. In 1D chains
with frustrated exchange interactions one--spin correlations
decrease exponentially with the distance from spin to spin. In
contrast pair, triple, and multiple correlations of adjacent spins
within a chain can decrease with power--law behavior. In this case
the phase is named the quasi--long range ordered phase. An external
static magnetic field $H$ or a magnetic anisotropy in such 1D spin
systems can establish true long--range order with multi--spin tensor
order parameter.\cite{Kolezhuk_05, Hikihara_08, Sudan_09,
Heidrich_09} It is a compelling task to find this magnetic state
with tensor order parameter in real magnetic materials. There are
plausible statements about possible magnetic phases in frustrated
chain magnets: the first statement is that interchain interaction in
real materials can cause true long--range order instead of quasi
long--range order in the 1D case. For instance, quasi long--range
ordered in the case of 1D the so called spin--nematic phase starts
to be long--range ordered in the presence of small ferromagnetic
interchain exchange interaction for the two--dimensional (2D) case,
as it was concluded theoretically in Refs.(\onlinecite{Ueda_09,
Zhitomirsky_10, Nishimoto_10}). The second statement is that spin
fluctuations can be strongly depressed by interchain interactions,
and conventional magnetic order with non--zero mean value of the
magnetization at the magnetic ions occurs in the particular magnetic
material under investigation. These serious restrictions give rise
to the fact that tensor magnets are not observed experimentally yet.

The subject of our report is a high--field NMR study of the
frustrated quasi one--dimensional $S=1/2$ chain cuprate LiCuVO$_4$.
In this compound, magnetic frustration is due to the intrachain NN
ferromagnetic and the NNN antiferromagnetic exchange. In small
magnetic fields $H$ and temperatures $T<T_{\rm N}$ (with $T_{\rm
N}\approx 2.3 $~K) an incommensurate planar spiral structure of the
magnetic Cu${}^{2+}$ moments is realized. The orientation of the
spin plane is defined by the direction and value of the applied
field $H$ and the crystal anisotropy. \cite{Buettgen_07} The wave
vector $\mathbf{k_{ic}}$ of this structure is directed along the
chains and is defined by intrachain exchange parameters $J_{\rm NN}$
and $J_{\rm NNN}$(cf. Ref. \onlinecite{Enderle_05}). In higher
fields $H
> H_{\rm c2}$ (with $\mu_0H_{\rm c2} \approx 7.5$~T) the collinear,
spin--modulated structure is realized. In this phase spins are
collinear with the direction of the applied field $H$ and their
values are modulated along the chain.\cite{Buettgen_07,Buettgen_10}
If the spiral structure can be described in the semiclassical
approach,\cite{Nagamiya_62} the spin--modulated phase starts to be
preferable due to quantum and thermal fluctuations. The theoretical
analysis of frustrated spin chains with the intrachain exchange
parameters of LiCuVO$_4$ in the low--field range revealed a
long--range ordered spin--chiral phase at $T = 0$. For higher fields
the spin--density wave (SDW) phase is expected,\cite{Hikihara_08,
Sudan_09,Heidrich_09} which is quasi long--range ordered. From the
experimental point of view the low--field magnetic phases in
LiCuVO$_4$ are long--range ordered and are characterized by magnetic
correlations similar to that of 1D theory. In contrast to the 1D
case the spiral and spin--modulated magnetic structures of
LiCuVO$_4$ are pinned, i.e., the values of the magnetic moments at
the Cu$^{2+}$ ions are non zero in LiCuVO$_4$, but strongly reduced
due to spin fluctuations. Note that a SDW phase is not peculiar only
in 1D frustrated systems. A long--ranged ordered SDW phase recently
was detected theoretically in a 2D spin system ($S$ = 1/2) with
distorted triangular lattice.\cite{Starykh_2006, Starykh_2010}

By further increase of the applied field $H$ in the 1D case it is
expected that the spin--nematic phase develops just before the
magnetically saturated phase occurs at the end. In recent
experiments the magnetization curve $M(H)$ of LiCuVO$_4$ exhibited
anomalies just before the saturation field $H_{\rm sat}$
corresponding to a new magnetic phase. \cite{Svistov_11} Probably,
this phase is the spin--nematic phase with zero mean value of the
magnetic moments at the Cu$^{2+}$ ions. In our previous NMR studies
we applied magnetic fields $H$ up to $\mu_0H=12$~T (Refs.
\onlinecite{Buettgen_07, Buettgen_10}). From those experiments we
were able to conclude that for $H < H_{c2}$ the spiral spin
structure is realized. Here we show that for higher fields
$H_{c2}<H$ the spin--modulated magnetic structure is established at
least up to $\mu_0H=30$~T.

In Sec. \ref{section2} we present a short review of the status quo
in LiCuVO$_4$ concerning the crystallographic and magnetic
structures of this compound. Sec. \ref{section3} is devoted to the
description of experimental techniques. There are two parts of
experimental results in Sec. \ref{section4}: in the first are
presented the NMR spectra obtained at different frequencies and
temperatures,  and in the second there are described relaxation
properties of nuclear spins. Using the values of hyperfine constants
for $^7$Li and $^{51}$V obtained experimentally in high--field
experiments we simulated the NMR spectra for different models of
magnetic structures possibly realized in LiCuVO$_4$. According to
the present theoretical concepts the collinear spin--modulated
structure in LiCuVO$_4$ constitutes a SDW phase which was predicted
for 1D frustrated $S=1/2$ systems.\cite{Hikihara_08,
Sudan_09,Heidrich_09} Taking into account that the strongest
exchange interactions in LiCuVO$_4$ are intrachain interactions we
expect that the wave vector deduced from experiments has the same
field dependence as in the 1D case.\cite{Maeshima_05, Hikihara_08}
This assumption enables us to reconstruct the magnetic field
dependence of the spin--modulated structure from our NMR spectra
within the range $H_{c2}<H<\mu_0H=30$~T as documented in section
\ref{section5}.

\section{CRYSTALLOGRAPHIC AND MAGNETIC STRUCTURES}
\label{section2}

\begin{figure}
\includegraphics[width=60 mm,angle=0,clip]{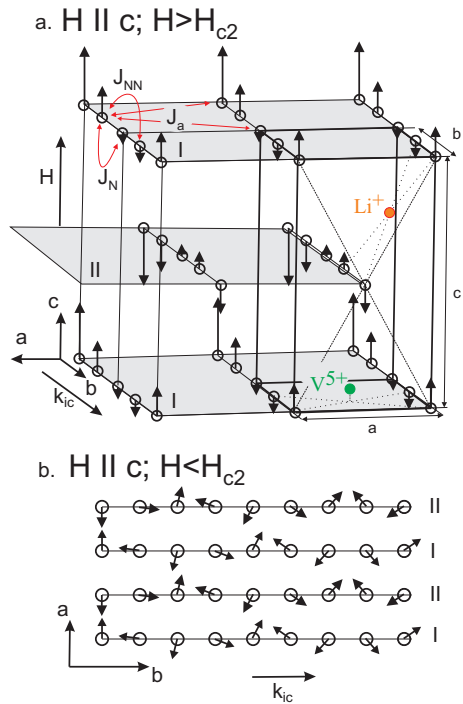}
\caption{(a) Scheme of the Cu$^{2+}$ moments in the crystal
structure of LiCuVO$_4$. Copper ions are marked by open circles with
arrows at each site which constitute the spin modulated structure
for $H>H_{c2}$, $\mathbf{H}\parallel \mathbf{c}$, and $T<T_{\rm N}$.
The red arrows $J_{\rm N},J_{\rm NN},J_{a}$ denote the main exchange
integrals (Ref. \onlinecite{Enderle_05}). Additionally, the
positions of Li (orange) and V (green) ions are exemplarily depicted
by one ion each. (b) Projection of the magnetic structure of two
{\bf ab}--planes for $H<H_{c2}$ and $\mathbf{H}\parallel
\mathbf{c}$.} \label{fig-1}
\end{figure}

\begin{figure}
\includegraphics[width=80 mm]{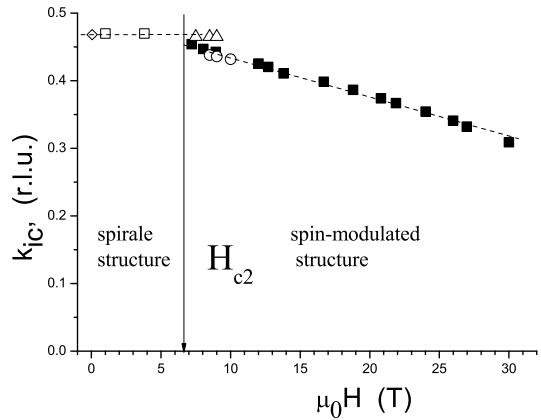}
\caption{Field dependence of the incommensurate wave vector $k_{\rm
ic}$ for applied magnetic fields $\mathbf{H}\parallel \mathbf{c}$ in
LiCuVO$_4$. The open symbols are data from neutron diffraction of
different authors: ($\diamond$, Ref. \onlinecite{Gibson_04}),
($\Box$, Ref. \onlinecite{Mourigal_10}), ($\triangle$ and $\circ$,
Ref. \onlinecite{Masuda_11}). The solid squares are values obtained
from magnetization measurements (Ref. \onlinecite{Svistov_11}) in
the frame work of the 1D model using Eq. (\ref{eq:3}).}
 \label{fig-2}
\end{figure}

LiCuVO$_4$ crystallizes in an inverse spinel structure $AB_2$O$_4$
with an orthorhombic distortion induced by a cooperative
Jahn--Teller effect of the Cu$^{2+}$ ions at octahedral sites. The
crystal structure belongs to the space group of symmetry {\em Imma}.
The elementary cell contains four magnetic ions Cu$^{2+}$($S=1/2$)
with the coordinates (0,0,0), (0,1/2,0), (1/2,0,1/2), and
(1/2,1/2,1/2) (see Fig. \ref{fig-1}a). The first two ions lie in the
{\bf ab}--plane which is marked as I, and the latter two in the {\bf
ab}--plane II. From elastic neutron--diffraction experiments it was
established \cite{Gibson_04} that in the low--temperature phase for
$T < T_N$ and zero applied magnetic field $H=0$ an incommensurate
planar spiral spin structure forms which has the propagation wave
vector $\mathbf{k_{ic}}$ directed along the Cu$^{2+}$ chains
($\mathbf{k_{ic}}\parallel\mathbf{b}$) (Fig. \ref{fig-1}b). We
parametrized the spiral of this spin structure with magnetic moments
$\mu_{\rm Cu}$ of the Cu$^{2+}$ ions utilizing the coordinates $x$,
$y$, and $z$ along the $\mathbf{a}$,$\mathbf{b}$, and $\mathbf{c}$
directions, respectively (Ref. \onlinecite{Buettgen_07}):

\begin{eqnarray}
\label{eq:1} \mathbf{\mu}(x,y,z)=\mu_{\rm Cu} \cdot \cos(2\pi
z/c)\cdot [ \mathbf{l_1} \cdot \cos(k_{ic}\cdot y+\phi)+
\\ \nonumber \mathbf{l_2} \cdot \sin(k_{ic}\cdot
y+\phi)]
\end{eqnarray}

where  $\mathbf{l_1}$ and $\mathbf{l_2}$ are orthogonal unit vectors
within the {\bf ab}--plane. At zero applied magnetic field $H=0$,
the absolute value of the propagation wave vector is k$_{\rm ic}$ =
$(1-0.532) \cdot 2\pi/b$ and the ordered Cu$^{2+}$ moment amounts to
$\mu_{\rm Cu}=0.31\mu_B$ (Refs. \onlinecite{Gibson_04,Enderle_05})
or $\mu_{\rm Cu}=0.25 \mu_B$ (Ref. \onlinecite{Yasui_08}). The angle
$\phi$ in Eq. (\ref{eq:1}) denotes an arbitrary phase shift. Figure
\ref{fig-1}b shows a stacking sequence of the above mentioned {\bf
ab}--planes I and II of this structure projected along the {\bf c}
--axis. The mutual orientation of spins of neighboring chains is the
following: the nearest spins of neighboring chains within the same
{\bf ab}--plane are parallel, whereas nearest spins of neighboring
chains from adjacent {\bf ab}--planes (planes I and II in Fig.
\ref{fig-1}) are antiparallel. The magnetic susceptibility $\chi$ of
this structure exhibits an anisotropy with slightly higher values
for magnetic fields applied perpendicular to the spin plane compared
to the values with the magnetic field applied within the spin plane.
This fact explains the spin--flop transition of reorienting magnetic
moments, when an applied magnetic field $\mu_0 H_{c1}\approx 2.5 $~T
is applied within the {\bf ab}--plane of the
crystal.\cite{Buettgen_07} According to our results of NMR
measurements the effective magnetic moment of the Cu$^{2+}$ ions
within the entire field range of the incommensurate planar spiral
structure is nearly constant $\mu_{\rm Cu}=0.3\mu_B$ (Ref.
\onlinecite{Buettgen_07}). Recent results of neutron--scattering
experiments\cite{Mourigal_10} show that with experimental accuracy
the incommensurate wave vector $k_{\rm ic}$ does not change within
this phase as it is displayed in figure \ref{fig-2}.

A more interesting and unexpected phase transition is observed at
elevated magnetic fields $\mu_0H_{c2}\approx 7.5$~T. The observation
of this magnetic transition for all three directions $\mathbf{H}
\parallel \mathbf{a},\mathbf{b},\mathbf{c}$ reveals an exchange nature of
this transition. The NMR spectra observed at $H > H_{c2}$ can be
well explained by the assumption that a collinear spin--modulated
structure is realized.\cite{Buettgen_07} The scheme of this
structure is shown in Fig. \ref{fig-1}a and can be parametrized as:

\begin{eqnarray}\label{eq:2}
\mathbf{\mu}(x,y,z)= \mathbf{l} \cdot [\mu_{\rm m}+\mu_1 \cdot
\cos(k_{ic}\cdot y+\phi_{I,II})],
\end{eqnarray}

where the magnetic moments of Cu$^{2+}$ ions $\mu$ are parallel to
the applied magnetic field $H$, i.e., the unit vector $\mathbf{l}
\parallel \mathbf{H}$. Moreover, $\mu_{\rm m}$ is the magnetization of the
sample per single Cu$^{2+}$ ion and $\mu_1(H)$ is the modulation
strength of the structure. The angles $\phi_{I}$ and $\phi_{II}$
denote phase shifts within the particular {\bf ab}--planes indexed
$I$ and $II$ (cf. Fig. \ref{fig-1}a), where
$\phi_{II}=\phi_{I}+\pi$.

Figure \ref{fig-2} summarizes all values of the incommensurate wave
vector $k_{ic}$ of the spin modulated structure as a function of the
applied magnetic field $H$ as obtained from literature. Up to
$\mu_0H=10$~T the data are taken from Refs.
\onlinecite{Gibson_04,Mourigal_10,Masuda_11}. Additionally, the
values of $k_{\rm ic}(H)$ (solid squares) obtained for the 1D model
as extracted from magnetization $M(H)$ measurements\cite{Svistov_11}
of LiCuVO$_4$ are plotted. In the frame work of this model the field
dependence of $k_{\rm ic}$ is expected to
be:\cite{Maeshima_05,Hikihara_08}

\begin{eqnarray}\label{eq:3}
k_{ic}(H)=(1-\frac{\mu_{\rm m}}{g\mu_BS})\cdot\frac{\pi}{b}
\end{eqnarray}
The experimental values $k_{\rm ic}(H)$ are in good agreement with
each other and fit with the 1D model according to equation
(\ref{eq:3}). Therefore, for the simulation of our NMR spectra of
LiCuVO$_4$ in this work we use the field dependence of the
incommensurate vector defined by Eq. (\ref{eq:3}) within the entire
magnetic field range under investigation. As it was elaborated in
Ref. \onlinecite{Buettgen_10} the spin--modulated phase exhibits
long--range magnetic order of Cu$^{2+}$ moments only within separate
{\bf ab}--planes, whereas the antiferromagnetic order in {\bf
c}--direction is only short--ranged.

\section{SAMPLE PREPARATION AND EXPERIMENTAL DETAILS}
\label{section3}

Single crystals of several cubic millimeters were grown as described
in Ref. \onlinecite{Prokofiev_05}. The single crystalline samples
studied in the present work satisfy the following stoichiometry
conditions: Li/V=0.96$\pm$0.05, Li/Cu=0.95$\pm$0.04,
Cu/V=0.99$\pm$0.01. Thus the average composition of the sample is
Li$_{0.97}$CuVO$_4$. The analysis of the sample quality is discussed
in Refs. \onlinecite{Prokofiev_04,Svistov_11}. The single crystal
used in the present work is the identical single crystal which was
studied in our previous low--field NMR experiments
\cite{Buettgen_07, Buettgen_10, Svistov_11} and was denoted in Ref.
\onlinecite{Prokofiev_04} as sample from the batch 1.

\begin{figure}
\includegraphics[width=80 mm]{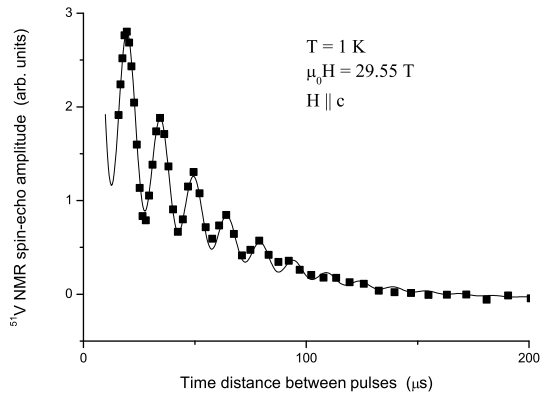}
\caption{Time dependence of the integral amplitude $M_{x,y}(2\tau)$
of the spin--echo signal (symbols) for $\mathbf{H}\parallel
\mathbf{c}$, $\mu_0H=29.55$ T, $\nu=336$ MHz at $T=1$~K. Solid line:
result of fitting with Eq. (\ref{eq:4}).}
 \label{fig-3}
\end{figure}

The NMR experiments were performed with a phase coherent, homemade
spectrometer at radio frequencies within the range $70<\nu<340$~MHz
at the National High Magnetic Field Laboratory, Tallahassee, USA. We
investigated the $^{7}$Li ($I$=3/2, $\gamma$/2$\pi$=16.5466 MHz/T)
and $^{51}$V ($I$=7/2, $\gamma$/2$\pi$=11.2133 MHz/T) nuclei using
spin--echo techniques with a pulse sequence
3$\mu$s--$\tau$--3$\mu$s. All NMR spectra were collected with the
pulse separation $\tau $= 15 $\mu$s by sweeping the applied magnetic
field $H$ within $7<\mu_0H<30$~T at constant frequencies, and the
temperatures were stabilized with a precision better than 0.02~K.
Fig. \ref{fig-3} shows the decay of the $^{51}$V integral amplitude
of the spin--echo signal $M_{x,y}$ on varying $\tau$ (the
conventional $T_2$ experiment). The time dependence exhibits an
exponential decay superimposed by pronounced oscillations which can
be fitted with:\cite{Degani_72}

\begin{eqnarray}\label{eq:4}
M_{x,y}(2\tau)=exp(-\frac{2\tau}{T_2})\cdot[C_0+C_1\cos({2a\tau+\delta})],
\end{eqnarray}
where $C_0$, $C_1$ and $\delta$ are dimensionless constants, and the
oscillations frequency $a$ has the value of the quadrupole frequency
for LiCuVO$_4$. The origin of the oscillations due to the
interaction between the nuclear quadrupole moment and a local
electric field gradient was discussed in reference
\onlinecite{Kegler_06}.

\section{EXPERIMENTAL RESULTS}
\label{section4}

\begin{figure}
\includegraphics[width=80 mm]{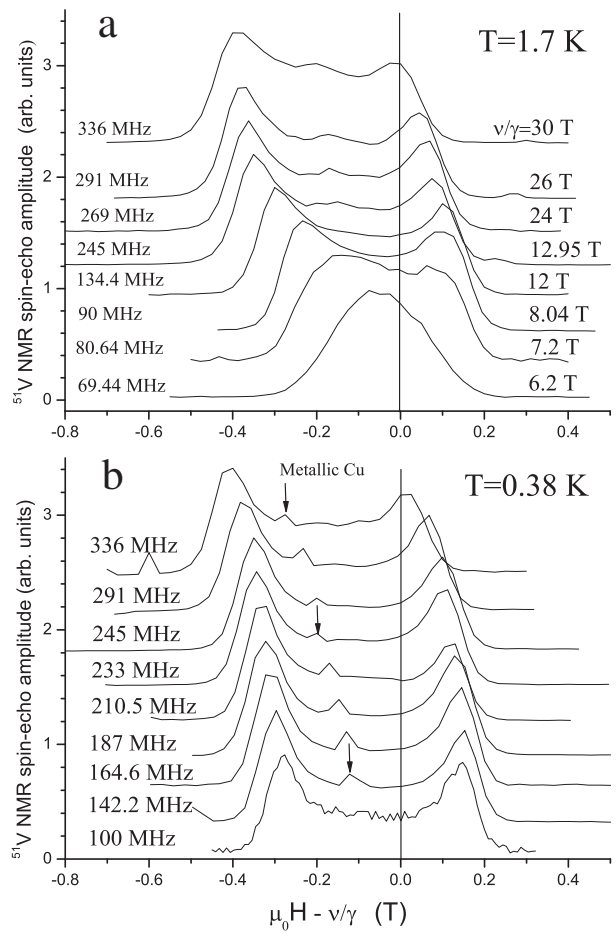}
\caption{$^{51}$V NMR spectra at $T = 1.7$ and 0.38 K for
$\mathbf{H}\parallel \mathbf{c}$ (upper and lower panel,
respectively). Residual signals from metallic copper within the
experimental setup are marked with arrows.
 }
 \label{fig-4}
\end{figure}

\begin{figure}
\includegraphics[width=80 mm]{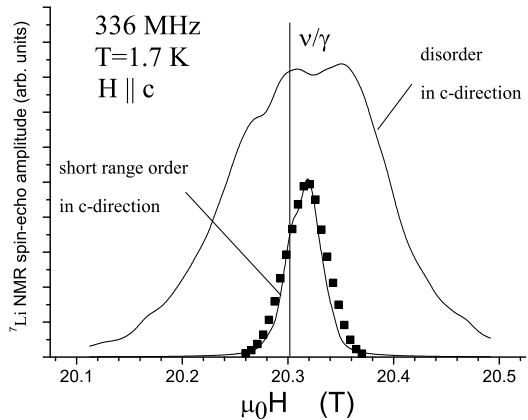}
\caption{$^{7}$Li NMR spectrum (black squares) at 1.7 K ($T<T_N$)
for $\nu=336$~MHz and $\mathbf{H}\parallel \mathbf{c}$. The solid
lines represent simulations of the NMR spectrum: lower curve was
computed in assumption of short--range order in {\bf c}--direction,
and the upper curve for disorder in {\bf c}--direction.
 }
 \label{fig-5}
\end{figure}

Figures \ref{fig-4}a and b show the NMR spectra of the $^{51}$V
nuclei obtained in the magnetically ordered phase at $T = 1.7$ and
0.38 K below the Neel temperature $T_{\rm N}$, respectively. The
spin--echo spectra exhibit a single line for small applied magnetic
fields $H < H_{\rm c2}$ and a double--horn shaped pattern for higher
fields. The spectra obtained at fields less than $\mu_0H<12$~T are
in a good agreement with previous
results.\cite{Buettgen_10,Smith_2006} The line shape of the NMR
spectra do not change between $10 < \mu_0 H < 30$~T which indicates
that the spin--modulated magnetic structure established in Refs.
\onlinecite{Buettgen_07,Buettgen_10} is robust within this extended
field range. The same consideration accounts for the case of
$^{7}$Li NMR spectra as the un--split, single--line
pattern\cite{Buettgen_07, Buettgen_10} for $H>H_{c2}$ is maintained
up to $\mu_0H = 20$~T (see Fig. \ref{fig-5}, black squares).

\begin{figure}
\includegraphics[width=80 mm]{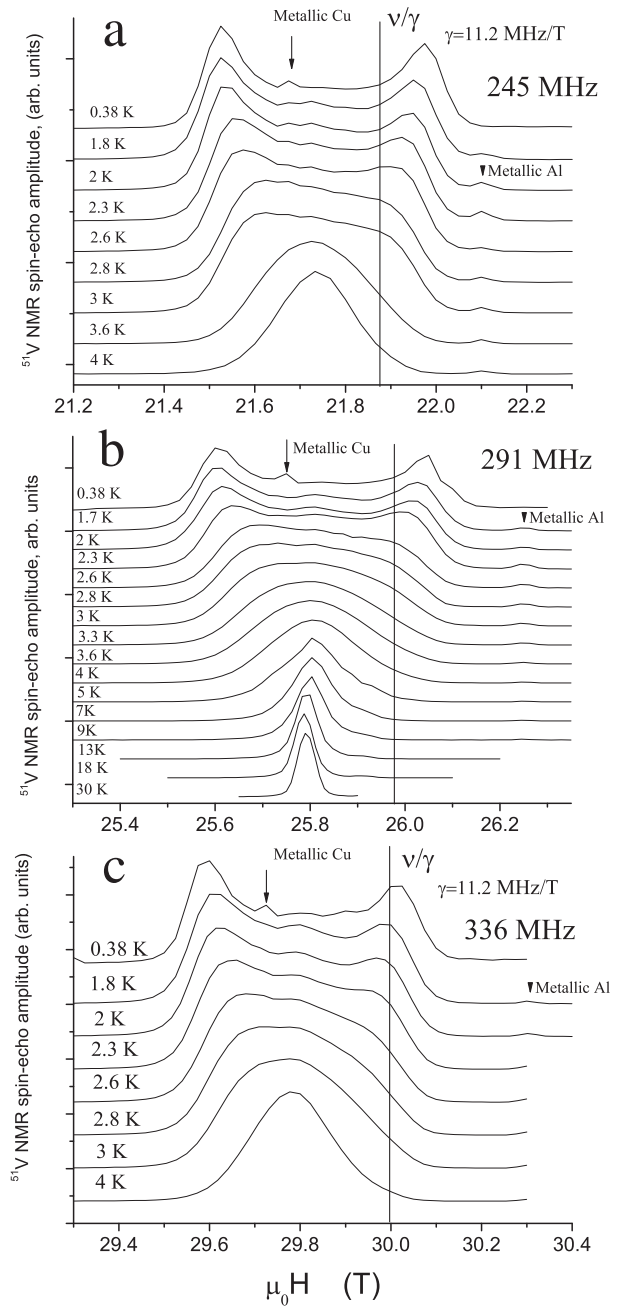}
\caption{Temperature dependence of $^{51}$V NMR spectra obtained  at
$\nu =$245, 291, and 336 MHz (panels a, b, and c) for
$\mathbf{H}\parallel \mathbf{c}$. Residual signals from metallic
aluminum within the experimental setup are marked with arrows.
 }
 \label{fig-6}
\end{figure}

Small sharp resonance anomalies due to metallic copper and aluminum
of the experimental setup are indicated by arrows as artefacts in
figures \ref{fig-4} and \ref{fig-6}.

In order to trace the field dependence of $T_{\rm N}$ the
temperature evolution of the NMR spectra was studied at elevated
magnetic fields $\mu_0 H = 22$, $26$, and $30$~T (at corresponding
NMR frequencies 245, 291, and 336 MHz, respectively). These
measurements are given in Fig. \ref{fig-6} and it is observed that
the transition from a single line in the paramagnetic regime to a
double--horn pattern deep in the spin--modulated phase below $T_{\rm
N}$ is not very sharp. The development of the double--horn pattern
toward low temperatures is accompanied by an additional unshifted
spectral line at the same field of the paramagnetic signal, i.e.,
the paramagnetic signal survives in the magnetically long--range
ordered phase at least down to temperatures around 1.7 K. Hindered
due to this gradual behavior we obtain the transition temperature
$T_{\rm N}$  by integrating the area of the spectra pattern. The
values of $T_N$ for different fields $H$, which are plotted in the
$(H,T)$ magnetic phase diagram (see Fig.\ref{fig-11}), were obtained
with this procedure.

\begin{figure}
\includegraphics[width=80 mm]{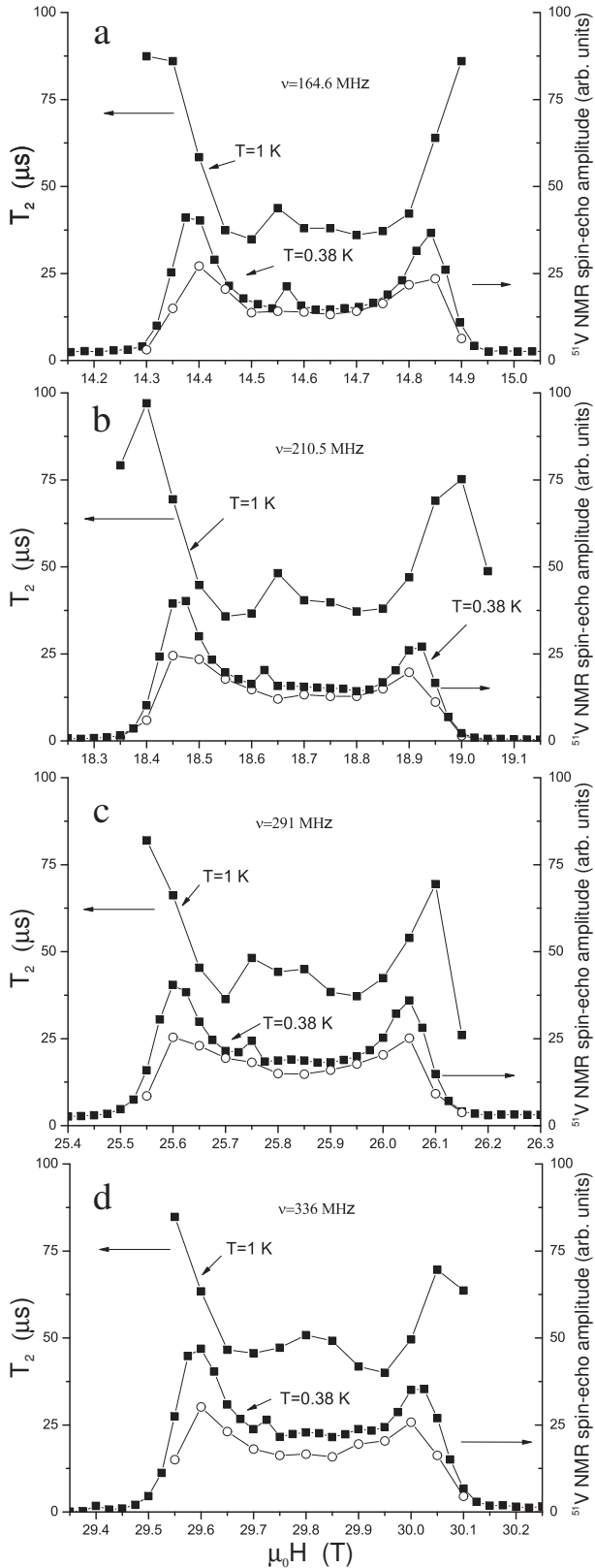}
\caption{$^{51}$V NMR spectra and field dependences of the
spin--spin relaxation time $T_2$ measured at frequencies 164.6 MHz
(a), 210.5 MHz (b), 291 MHz (c), and 336 MHz (d) for
$\mathbf{H}\parallel \mathbf{c}$ in LiCuVO$_4$ (solid squares).
Spectra were measured at $T=0.38$~K and $T_2$ values at $T=1$~K,
respectively. Open circles correspond to the expected NMR spectra
after a so called $T_2$--correction (see text).
 }
 \label{fig-7}
\end{figure}

The spin--spin relaxation time $T_2$ of $^{51}$V depends on the
particular position of the probing nuclei within the magnetically
ordered structure. The field dependence of T$_2$ for different
frequencies is shown in Fig. \ref{fig-7} (left axis). In the same
figure the NMR spectra with their characteristic double--horn shaped
pattern are presented as well. The $T_2$ measurements were performed
at $T=1$~K and the spectra were measured at $T=0.38$~K. From Fig.
\ref{fig-6} it is clear that the NMR spectra of $^{51}$V nuclei do
not change significantly within the temperature range $0.38 < T <
1$~K. This allows us to analyze these data together. $T_2$
monotonically growths toward the borders of the spectra. It is
important to note that the maxima of $T_2$ occur at applied magnetic
field values $H$ which do not coincide with the particular positions
of the low-- and high--field maxima of the double--horn spectra
pattern.

\section{Discussion and Conclusion}
\label{section5}

\begin{figure}
\includegraphics[width=80 mm]{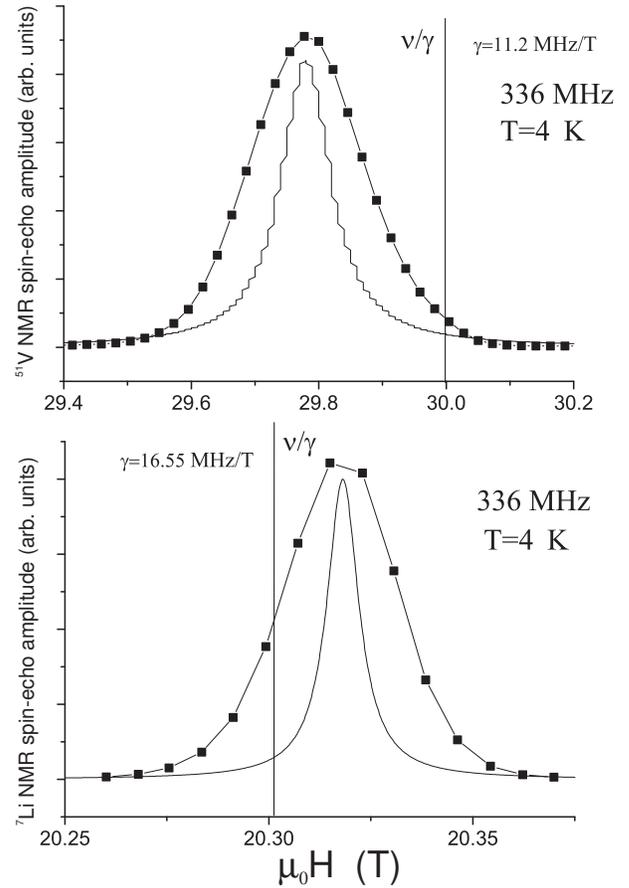}
\caption{$^{51}$V and $^7$Li NMR spectra in the paramagnetic phase
($T=4$ K) for $\nu= 336$ MHz and $\mathbf{H}\parallel \mathbf{c}$.
The solid lines show the NMR spectra simulations with hyperfine
constants of the contact interaction given in the text.
 }
 \label{fig-8}
\end{figure}

The effective magnetic field at the $^7$Li or $^{57}$V nuclei in
LiCuVO$_4$ is defined by three items $\mathbf{H_{\rm
eff}}=\mathbf{H}+\mathbf{H_{\rm dip}}+\mathbf{H_{\rm cont}}$, i.e.,
the externally applied field $\mathbf{H}$, the dipolar long--range
field from magnetic neighbors $\mathbf{H_{\rm dip}}$, and the
short--range Fermi--contact field $\mathbf{H_{\rm cont}}$ induced by
nearest neighboring magnetic moments. The $^7$Li and $^{51}$V nuclei
have four (N$=4$) nearest magnetic neighbors, which are highlighted
in Fig. \ref{fig-1}a with dotted lines. For $^7$Li nuclei the
nearest magnetic neighbors are from different {\bf ab}--planes (I
and II, respectively), whereas for $^{57}$V nuclei the nearest
magnetic neighbors are located within the same {\bf ab}--planes (I
or II, see Fig. \ref{fig-1}a). All copper sites in LiCuVO$_4$ are
crystallographically equivalent. This allows to define the contact
fields at the nuclei under investigation (both $^7$Li and $^{57}$V)
by the magnetic moments of their four nearest neighboring moments
$\mu_i$ as $\mathbf{H_{\rm
cont}}=A(\mathbf{\mu_1}+\mathbf{\mu_2}+\mathbf{\mu_3}+\mathbf{\mu_4})$,
where the hyperfine coupling $A$ is a second--rank tensor. Its
values $A_{zz}$ were obtained from the NMR study in the paramagnetic
phase. In the paramagnetic phase each magnetic ion has the same
magnetic moment $\mu_{i,\rm z}=M_{\rm z}/N$, where $M_{\rm z}$ is
the magnetic moment of the sample measured in an applied magnetic
field $\mathbf{H}\parallel \mathbf{c}$, and $N$ is the number of
magnetic ions. We took the value of $M_{\rm z}$ in an applied field
from Ref. \onlinecite{Svistov_11}. The dipolar field $\mathbf{H_{\rm
dip}}$ we computed numerically and $\mathbf{H_{\rm dip}}$ is
determined in the paramagnetic phase by only one single parameter,
the magnetic moment of each magnetic ion $\mu_{i,\rm z}=M_{\rm
z}/N$. Thus there is enough information to deduce the values of the
hyperfine constants $A_{\rm zz}=(\nu/\gamma-H-H_{\rm dip,z})/(4
\cdot M_{\rm z}/N)$. The values of the hyperfine constants for
$^{51}$V and $^{7}$Li in LiCuVO$_4$ are $^{51}A_{\rm zz}= 0.125 \pm
0.01$~T/$\mu_0\mu_B$ and $^7A_{\rm zz}=-0.035\pm
0.01$~T/$\mu_0\mu_B$, respectively. The value of $^{51}A_{\rm zz}$
obtained in the high--field range is consistent with the value
obtained in the low--field range, whereas the value of $^{7}A_{\rm
zz}$ in the low--field range was evaluated as negligible
small.\cite{Buettgen_07} NMR spectra measured in the paramagnetic
phase at $T=4$ K are shown in figure \ref{fig-8}. The solid lines in
this figure represent NMR spectra simulations with the hyperfine
constants of the contact interaction $A_{\rm zz}$ as mentioned
above. The line widths of the simulated spectra were left as free
parameters.
The knowledge of the parameters $A_{\rm zz}$ allow to compare the
experimentally obtained NMR spectra with spectra expected
theoretically from different models. The amplitude of the NMR
spin--echo signal is given by the number of nuclei resonating at the
particular value of the effective field $H_{\rm eff}$, but also by
the spin--spin relaxation time $T_2$. In order to compensate for the
latter effect we multiplied the experimentally collected spectra by
the factor $\exp(2\tau/T_2)$(see Eq. (\ref{eq:4})), where the values
of $T_2$ have to be measured as a function of the applied magnetic
field $H$ (solid squares in Fig. \ref{fig-7}, left axis). This is
the so called $T_2$--correction of the spectral intensity. The
resulting spectra after $T_2$--correction are shown in Fig.
\ref{fig-7} (right axis) with open circles. We conclude that the
asymmetry of the uncorrected spectra (solid squares in Fig.
\ref{fig-7}, right axis) is a result of the field dependence of
$T_2$, whereas the positions of the characteristic maxima of the
double--horn pattern do not change after the $T_2$--correction.

\begin{figure}
\includegraphics[width=80 mm]{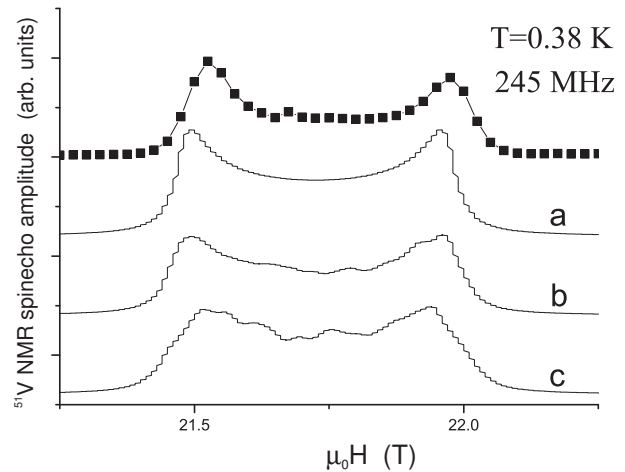}
\caption{$^{51}$V NMR spectrum in the high--field phase for $T =
0.38$~K $\ll T_{\rm N}$, $\nu=245$ MHz, and $\mathbf{H}\parallel
\mathbf{c}$ (solid squares). The solid lines show simulations of the
NMR spectrum utilizing the model of the spin--modulated phase (see
Eq. (\ref{eq:2})) with long--range order ('line a'), short--range
order ('line b'), and disorder ('line c') in {\bf c}--direction,
respectively.
 }
\label{fig-9}
\end{figure}

A closer inspection of the $T_2$--corrected $^{51}$V NMR spectrum at
$T = 0.38$~K and 245~MHz ($\mu_0H \approx 21.75$~T) is given in
figure \ref{fig-9}. Here, experimental data (solid squares) are
compared with three simulations based on three different models of
the magnetically ordered structure in the high--field phase of
LiCuVO$_4$ (solid lines a,b, and c in Fig. \ref{fig-9}). The model
of the magnetic structure for 'line a' is based on Eqs. (\ref{eq:2})
and (\ref{eq:3}). Here, the model includes a long--range order of
magnetic moments along the {\bf c}--direction. The 'line b' is
calculated supposing the phases $\phi_i$ in Eq. (\ref{eq:2}) to be
shifted by $\pi$ for two planes I and II nearest to the probing Li
and V nuclei (see Fig. \ref{fig-1}a), whereas the magnetic structure
of the other planes, which are more distant from the probing nuclei,
are also described by Eq. (\ref{eq:2}) but with random phase $\phi$.
This model of magnetic structure according to 'line b' we denote as
a structure with short--range order in {\bf c}--direction. The
underlying model of 'line c' emanates from random phases $\phi$
between all {\bf ab}--planes and the corresponding magnetic
structure is denoted as the magnetic phase with disorder in {\bf
c}--direction. In the framework of these models the only fitting
parameter was the modulation strength $\mu_1$ of the structure (see
Eq. (\ref{eq:2})). For the long--range and short--range ordered
magnetic structures ('line a and b') the values of $\mu_1$ at the
field of $\mu_0H \approx 21.75$~T were found to be equal to $0.8
\mu_B$. For the magnetic structure with disorder in {\bf
c}--direction ('line c') the value of the fitting parameter $\mu_1$
amounts to $0.65 \mu_B$. These fitting procedures with the three
different models mentioned above were applied to all measured
$^{51}$V NMR spectra collected in figure \ref{fig-4}.

\begin{figure}
\includegraphics[width=80 mm]{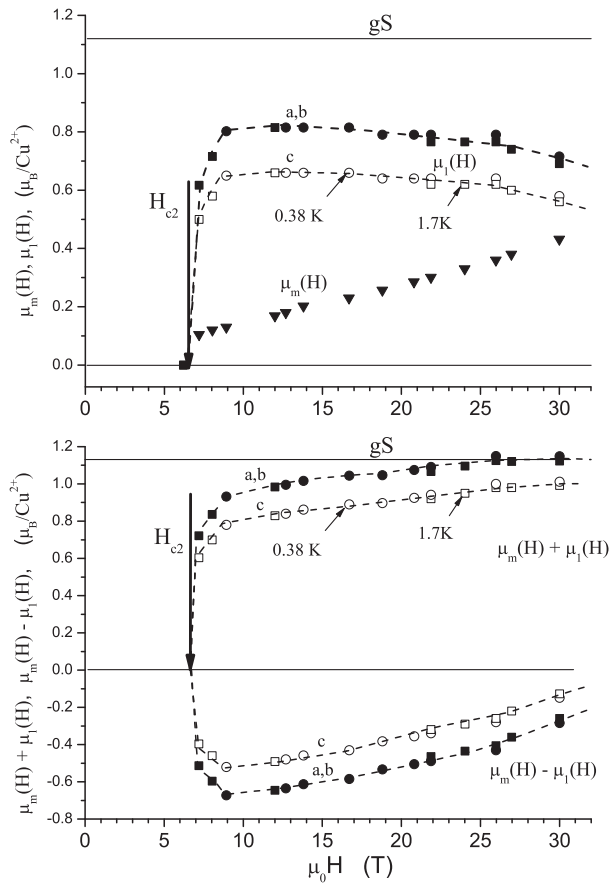}
\caption{Upper panel: field dependence of the magnetization
$\mu_{\rm m}$ and modulation strength $\mu_1$  of the
spin--modulated phase (see Eq. (\ref{eq:2})). The data $\mu_m(H)$
(triangles) are taken from Ref. \onlinecite{Svistov_11}. The values
$\mu_1(H)$ (circles and squares) are obtained from fitting NMR
spectra. Lower panel: field dependence of the extremal values of
magnetic moments in the spin--modulated structure as $\mu_m \pm
\mu_1$. For both panels solid symbols denote the fitting results
modeling long-- and short--range order in {\bf c}--direction. The
open circles and squares denote the fitting results modeling a
disorder in {\bf c}--direction. Squares and circles symbols
distinguish different temperatures of 1.7 and 0.38 K for
$\mathbf{H}\parallel \mathbf{c}$.
 }
 \label{fig-10}
\end{figure}

The field dependence of the modulation strength $\mu_1$ is shown in
the upper panel of figure \ref{fig-10}. The modulation strength
abruptly increases just above $H_{c2}$ and weakly decreases towards
higher applied magnetic fields $H$. The values of the ordered
magnetic moments of the Cu$^{2+}$ ions change along the chains in
the range between the extremal values $\mu_{\rm m} \pm \mu_1$ which
are plotted in the lower panel of figure \ref{fig-10}. The largest
value of the magnetic moment aligned along the applied magnetic
field $\mu_{\rm m} + \mu_1$ asymptotically yields the highest
possible value $gS\mu_B$ at highest magnetic fields $H$ (cf. lower
panel of Fig. \ref{fig-10}). Additionally, it is important to note
that the obtained values of $\mu_1$ in case of long--range order and
short--range order (i.e., model calculations 'line a' and 'line b'
in Fig. \ref{fig-9}, respectively) are almost the same. The third
simulated magnetic structure modeled with disorder in {\bf
c}--direction (cf. 'line c') yields values of $\mu_1$ which are
around 20$\%$ smaller than the former ones. As the effective field
at the probing $^{51}$V nuclei is predominantly given by
Fermi--contact fields $\mathbf{H_{\rm cont}}$ of Cu$^{2+}$ ions from
the nearest {\bf ab}--plane a weak sensitivity of $\mu_1$ to the
three different models of the magnetic structure is expected.

In contrast to the case of $^{51}$V NMR, the NMR spectra of $^7$Li
are very susceptible to the magnetic order in {\bf c}--direction.
Here, the probing lithium nuclei are situated between adjacent
planes I and II as shown in Fig. \ref{fig-1}a, and the effective
field predominantly depends on mutual orientations of spins from
these neighboring planes. Simulating the case of long--range
magnetic order using Eq. (\ref{eq:2}) straightforwardly accounts for
the single $^7$Li NMR spectral line determined by the applied field
$H$ and the magnetization $\mu_{\rm m}$ per Cu$^{2+}$ ion of the
sample. Corresponding simulations of this single $^7$Li NMR spectral
line for magnetic structures with short--range order and with
disorder in {\bf c}--direction, respectively, are shown in Fig.
\ref{fig-5}. Modeling the magnetic structure with short--range order
in {\bf c}--direction describes the observed NMR spectrum of $^7$Li
nuclei at the best. Note that in our previous low--field study
\cite{Buettgen_10} of the spin--modulated phase we preferred the
magnetic structure with disorder in {\bf c}--direction. However, at
that time our choice was strongly based on the assumption of
negligible Fermi--contact fields at the $^7$Li nuclei. Our new
high--field spectra document a significant contribution of
Fermi--contact fields acting on the lithium sites which allows us to
rule out our earlier model given in Ref. \onlinecite{Buettgen_10}.
Hence, the field dependences of $\mu_1(H)$ and $\mu_{\rm m} \pm
\mu_1(H)$ (see Fig. \ref{fig-10}, solid symbols) obtained from our
NMR spectra simulation within the model of short--range ordering
along the {\bf c}--direction are more realistic.

Our measurements of the spin--spin relaxation time $T_2$ at four
different NMR frequencies $\omega/2\pi = 165, 211, 291$, and
336~MHz, accordingly four different applied magnetic fields, reveal
two main results. Firstly, in the middle range of the field swept
$^{51}$V NMR spectra the value of $T_2$ is significantly shorter
than the values of $T_2$ at the maximas fields of the double--horn
shaped spectra pattern. Secondly, the relaxation times $T_2$ at the
low--field maxima of the NMR spectra pattern are systematically
longer than $T_2$ measured at the high--field maxima of the spectra.
This effect is more pronounced for elevated frequencies
$\omega/2\pi$. These two results are qualitatively explained in the
framework of the spin--modulated magnetic structure if we assume
that the spin--spin relaxation time at the $^{51}$V nuclei is
defined by fluctuations of the perpendicular components of the
magnetic moments of nearest Cu$^{2+}$ ions. The slowest fluctuations
we expect at the copper ions with the largest module of its
component of magnetic moments aligned parallel to the applied static
field $H$, i.e., in the vicinity of the maxima of the double--horn
NMR spectra pattern. The parallel components of the magnetic moments
reach the extremal values $\mu_{\rm m} \pm \mu_1$, where the $+$
sign belongs to the low--field maximum and the $-$ sign to the
high--field maximum of the double--horn, respectively. The fastest
fluctuations of the perpendicular components of the magnetic moments
of nearest Cu$^{2+}$ ions are expected for copper ions with zero
magnetic component along the applied static field $H$. The NMR
signal from vanadium nuclei near the copper ions with zero magnetic
moment is observed at applied fields $H$ in the middle range of the
double-horn pattern.

\begin{figure}
\includegraphics[width=80 mm]{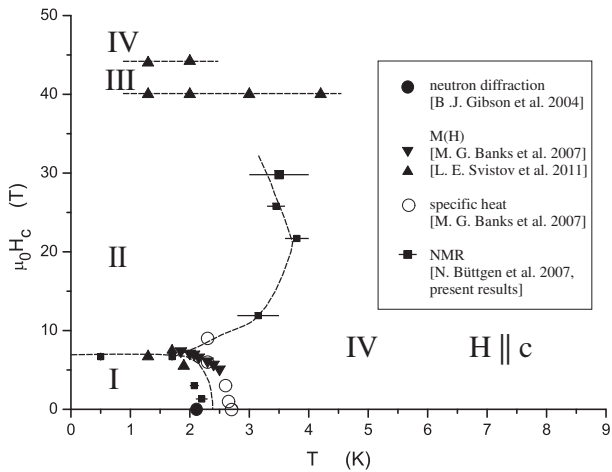}
\caption{$(H,T)$ phase diagram of LiCuVO$_4$ for
$\mathbf{H}\parallel \mathbf{c}$. The dashed lines are drawn to
guide the eye. Additional data are collected from Refs.
\onlinecite{Gibson_04, Banks_07,Buettgen_07, Svistov_11}.
 }
 \label{fig-11}
\end{figure}

Summarizing our high-field NMR experiments we present an extension
of the current $(H,T)$ magnetic phase diagram\cite{Gibson_04,
Banks_07,Buettgen_07, Svistov_11,Schrettle_2008} of LiCuVO$_4$ in
Fig. \ref{fig-11} for $\mathbf{H}\parallel \mathbf{c}$. This phase
diagram of LiCuVO$_4$ contains at least four phases: the planar
spiral phase (I), the collinear spin--modulated phase (II), possibly
the spin--nematic phase (III), and the polarized paramagnetic phase
(IV). The sequence of field dependent phase transitions at
temperatures $T\ll T_N$ is in qualitative agreement with results of
recent theoretical models for the 1D case (cf. Ref.
\onlinecite{Hagiwara_11}). As far as we know the complete $(H,T)$
magnetic phase diagram is theoretically not established up to now,
even for the 1D case  with frustrated intrachain exchange
interactions characteristic for LiCuVO$_4$.

In conclusion, we report on NMR studies of the quasi
one--dimensional (1D) antiferromagnetic $S=1/2$ chain cuprate
LiCuVO$_4$ in high--magnetic fields $H$ which amounts up to $\approx
70$~\% of the saturation field $H_{\rm sat}$. NMR spectra within the
field range $7.5 < \mu_0H < 30$~T for temperatures $T<T_{\rm N}$ can
be described within the model of a spin--modulated phase in which
the magnetic moments are aligned parallel to the applied field $H$
and their values alternate sinusoidally along the magnetic chains.
The measurements of the spin--spin relaxation time $T_2$ at the
vanadium sites reveal that $T_2$ depends on the particular position
of the probing $^{51}$V nucleus with respect to the magnetic copper
moments within the 1D chains: the largest $T_2$ value is observed
for the vanadium nuclei which are very next to the magnetic Cu$^{2+
}$ ion with largest ordered magnetic moment. This observation was
found in agreement with the expectation for the spin--modulated
magnetic structure in LiCuVO$_4$.

\begin{acknowledgments}
We thank O. A. Starykh and G. Teitel'baum for useful discussions.
This work is supported by the Grants 12-02-00557-\`{a},
10-02-01105-a of the Russian Foundation for Basic Research, Program
of Russian Scientific Schools, and by the German Research Society
(DFG) within the Transregional Collaborative Research Center (TRR
80).
\end{acknowledgments}

\end{document}